\documentclass[sigplan,nonacm]{acmart}

\usepackage{array}   
\usepackage{braket}
\usepackage{listings}
\usepackage{booktabs}
\usepackage{multirow}
\usepackage{graphicx} 
\usepackage{subcaption}  
\usepackage{xcolor}
\usepackage{tikz}
\usepackage{comment}

\definecolor{eprcolor}{RGB}{0,128,0}     
\definecolor{commcolor}{RGB}{0,0,255}    
\definecolor{gatecolor}{RGB}{128,0,128}  
\definecolor{meascolor}{RGB}{178,34,34}    
\definecolor{commentcolor}{RGB}{128,128,128} 
\definecolor{epr}{RGB}{255,165,0}   
\definecolor{quantum}{RGB}{139,69,19}       
\definecolor{classical}{RGB}{75,0,130}

\usepackage{enumitem}

\setlist[itemize]{leftmargin=*} 
\setlist[enumerate]{leftmargin=*} 

\author{Peiyi Li}
\affiliation{%
  \institution{North Carolina State University}
  \city{Raleigh}
  \state{North Carolina}
  \country{USA}
}
\affiliation{%
  \institution{Pacific Northwest National Lab }
  \city{Richland}
  \state{Washington}
  \country{USA}
}
\email{pli11@ncsu.edu}

\author{Chenxu Liu}
\affiliation{%
  \institution{Pacific Northwest National Lab}
  \city{Richland}
  \state{Washington}
  \country{USA}
}
\email{chenxu.liu@pnnl.gov}

\author{Ji Liu}
\affiliation{%
  \institution{Argonne National Laboratory}
  \city{Lemont}
  \state{Illinois}
  \country{USA}
}
\email{ji.liu@anl.gov}

\author{Huiyang Zhou}
\affiliation{%
  \institution{North Carolina State University}
  \city{Raleigh}
  \state{North Carolina}
  \country{USA}
}
\email{hzhou@ncsu.edu}

\author{Ang Li}
\affiliation{%
  \institution{Pacific Northwest National Lab}
  \city{Richland}
  \state{Washington}
  \country{USA}
}
\affiliation{%
  \institution{University of Washington}
  \city{Seattle}
  \state{Washington}
  \country{USA}
}
\email{ang.li@pnnl.gov}

\begin{document}
\title{A Quantum Network Processor Unit for Distributed Quantum Computing}

\begin{abstract}
As quantum computing progresses, the need for scalable solutions to address large-scale computational problems has become critical. Quantum supercomputers are the next upcoming frontier by enabling multiple quantum processors to collaborate effectively to solve large-scale computational problems. The emergence of quantum supercomputers necessitates an efficient interface to manage the quantum communication protocols between quantum processors. In this paper, we propose the Quantum Network Processing Unit (QNPU), which enables quantum applications to efficiently scale beyond the capacity of individual quantum processors, serving as a critical building block for future quantum supercomputers. The QNPU works alongside the Quantum Processing Unit (QPU) in our decoupled processing units architecture, where the QPU handles local quantum operations while the QNPU manages quantum communication between nodes. We design a comprehensive instruction set architecture (ISA) for the QNPU with high-level communication protocol abstractions, implemented via micro-operations that manage EPR resources, quantum operations, and classical communication. To facilitate programming, we introduce DistQASM, which extends OpenQASM with distributed quantum operations. We then propose a microarchitecture featuring both scalar and superscalar QNPU designs to enhance performance for communication-intensive quantum workloads. Finally, we evaluate the performance of our proposed QNPU design with distributed quantum workloads and demonstrate that the QNPU significantly improves the efficiency of communication between quantum nodes, paving the way for quantum supercomputing.
\end{abstract}

\maketitle 

\section{Introduction}\label{sec:intro}

As quantum computing advances, its potential to solve large-scale problems in domains such as cryptography~\cite{shor1994factor}, and chemistry problems~\cite{McArdle2020chemistry} becomes increasingly apparent. 
These large-scale problems often require computational resources far beyond the capacity of a single quantum processor.
One promising solution for large-scale problems is distributed quantum computing (DQC)~\cite{Cuomo2020DQC, Caleffi2024DQC}, which enables scalability by partitioning large quantum algorithms across multiple quantum processors, each contributing a portion of the overall computation.

Significant efforts have been devoted toward realizing DQC. Some works~\cite{Cuomo2020DQC, Rodrigo2020DQC, ang2024ARQUIN} propose full-stack layered architectures to capture the broad challenges of building distributed quantum systems. Others target specific challenges: At the software level, frameworks such as the Quantum Message Passing Interface (QMPI)~\cite{haner2021QMPI} extend the classical MPI standard to quantum systems, providing programming models for DQC. From the compiler level, a variety of frameworks~\cite{wu2022autocomm, wu2022collcomm, Wu2023QuComm} have been developed to optimize the scheduling of inter-node communication, thereby reducing the overhead of distributed execution. On the hardware side, demonstrations on various qubit technologies, including superconducting qubits~\cite{Magnard2020microwave, Smith2022Chiplet, Storz2023}, ion traps~\cite{Main2025dqc} have showcased the feasibility of distributed quantum systems.

Looking forward, the quantum computing community increasingly recognizes that the future of quantum computing lies not in a single technology, but in the strategic integration of diverse qubit types since no single qubit technology excels at all computational tasks. Recent initiatives such as the Defense Advanced Research Projects Agency (DARPA)'s Heterogeneous Architectures for Quantum (HARQ) program~\cite{DARPA_HARQ_2025} seeks to explore heterogeneous quantum systems to realize DQC. In addition, heterogeneous quantum architectures require specialized infrastructure capable of managing communication across different species of qubits.

Even within current homogeneous distributed quantum systems, existing architectures lack the specialized hardware needed for efficient inter-node communication. 
When distributed quantum algorithms require frequent communication between processors, existing approaches suffer from substantial performance bottlenecks due to the absence of dedicated communication processing units. This limitation becomes particularly severe for communication-intensive quantum algorithms, where the overhead of managing inter-node communications can dominate overall execution time.
For future heterogeneous systems, this challenge is further compounded by the need to transfer quantum states between qubits with different species.

\begin{figure}[h]
  \centering
  \includegraphics[width=\linewidth]{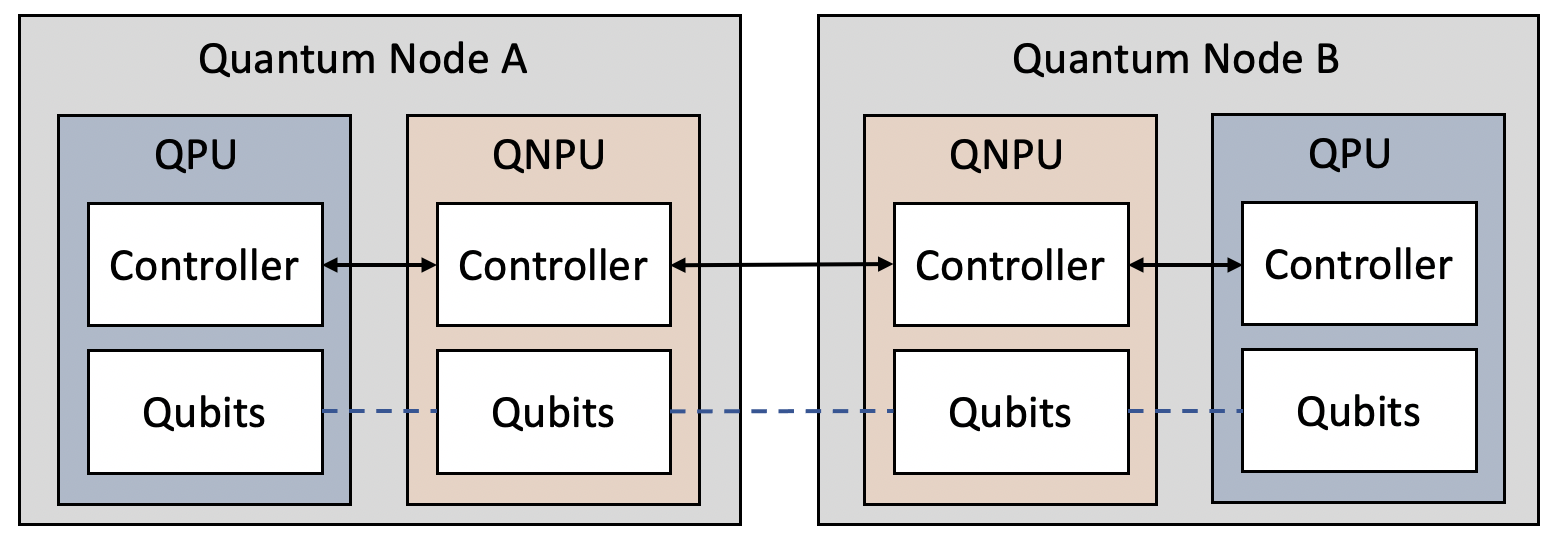}
\vspace*{-1\baselineskip}
  \caption{Organization of quantum nodes in a quantum supercomputer. Each quantum node integrates a Quantum Processing Unit (QPU) and a Quantum Network Processing Unit (QNPU). Both units contain a classical controller and a quantum chip (qubits). Black solid lines between classical controllers are classical links, and blue dashed lines between quantum chips are quantum links.
  While only two nodes are shown for clarity, a quantum supercomputer can scale to multiple nodes with similar interconnection patterns forming different topologies. }
  \label{fig:quantum_nodes}
\end{figure}

To support both homogeneous and heterogeneous distributed quantum computing systems, we introduce the concept of a quantum supercomputer: a scalable computing system composed of multiple quantum processors interconnected through quantum and classical links. At the core of this architecture is the Quantum Network Processor Unit (QNPU), a specialized hardware module designed to efficiently execute quantum communication protocols. By offloading communication tasks from the Quantum Processing Unit (QPU), the QNPU provides a clean separation between local computation and quantum communication, enabling efficient state transfer and coordination across processors. While our evaluation focuses on homogeneous systems, the same decoupled design naturally extends to heterogeneous quantum platforms. In such systems, different qubit technologies or hardware zones can interoperate through the QNPU, making it a foundational building block for future heterogeneous quantum computers.

As shown in Fig.~\ref{fig:quantum_nodes}, a quantum supercomputer is organized as interconnected quantum nodes. Each node has a Quantum Processor Unit (QPU) that executes local quantum computations within the quantum node, and a Quantum Network Processor Unit (QNPU) that performs quantum communication protocols between quantum nodes.
Both the QPU and QNPU consist of two key components: a classical controller managing instruction execution; and a quantum device containing physical qubits for quantum operations.
Within each quantum node, QPU and QNPU interact through both a classical link and a quantum link. Intra-node classical links connect their controllers, enabling the QPU to delegate quantum communication protocols to the QNPU for execution; Intra-node quantum links connect their quantum devices, enabling qubit movement between the QPU and the QNPU.
Communication between quantum nodes involves inter-node classical links and quantum links. Inter-node classical links support classical data transfer between nodes; inter-node quantum links support building the entanglement between nodes.

This decoupled architecture, which incorporates QPU and QNPU in one quantum node and separates local computation (handled by QPUs) from quantum communication protocols (managed by QNPUs), enables efficient execution of distributed quantum algorithms. By abstracting communication protocols away from the QPU implementation, this architecture also provides flexibility for future heterogeneous configurations where different nodes may employ different qubit technologies.

Our key contributions are as follows.
\begin{itemize}
\item Proposed a decoupled architecture for quantum supercomputers, which separates local computation (handled by QPUs) from quantum communication protocols (managed by QNPUs), providing a foundation for both  homogeneous and heterogeneous distributed quantum architectures.
\item Proposed a layered stack for quantum supercomputer architecture that breaks down the complexity of running distributed quantum algorithms.
\item Proposed DistQASM which extends OpenQASM to simplify programming for distributed quantum computing.
\item Proposed an instruction set architecture for QNPU to perform the quantum communication protocols between quantum nodes.
\item Designed the microarchitecture for QNPU for efficient execution of quantum communication protocols.
\item Implemented a cycle-level performance simulator to evaluate QNPU performance on distributed quantum workloads.
\end{itemize}

\section{Background and Related Work}\label{sec:background}
\subsection{EPR Pair Generation}
Inter-node quantum communication in a quantum supercomputer consumes entangled states between nodes for quantum state and gate teleportation~\cite{Wan2019,Chou2018, Hermans2022}. One of the widely used resource entangled states is Einstein–Podolsky–Rosen (EPR) pairs~\cite{Einstein1935epr}, represented as $\frac 1{\sqrt2}(\ket{00}+\ket{11})$. 

In contrast to quantum network applications, where EPR pairs are generated between remote nodes separated by long distances, typically on the order of kilometers~\cite{Wei2022, Knaut2024, Stolk2024}, we focus on distributed quantum algorithms running on the quantum supercomputers, where communication nodes are located within a relatively short range, such as within a single room. This proximity makes the generation of EPR pairs significantly less challenging and time-consuming compared to quantum networks.

For instance, in superconducting quantum computing systems, the nodes of a quantum supercomputer consist of superconducting chips~\cite{Smith2022Chiplet, guinn2023Quirc}. The physical connections enabling EPR pair generation can be coaxial cables that directly link superconducting qubits across different nodes—an approach that has already been experimentally demonstrated~\cite{Magnard2020microwave, Storz2023}. By cooling these coaxial cables to cryogenic temperatures, the loss of microwave photon propagation can be minimized. Consequently, EPR pairs between nodes can be efficiently generated through the direct exchange of microwave photons~\cite{Storz2023, Campagne-Ibarcq2018, Besse2020, Niu2023interconnect, Grebel2024}.

\subsection{EPR Pair Buffering}\label{subsec:epr_buffer}
\begin{figure}[h]
  \centering
  \includegraphics[width=\linewidth]{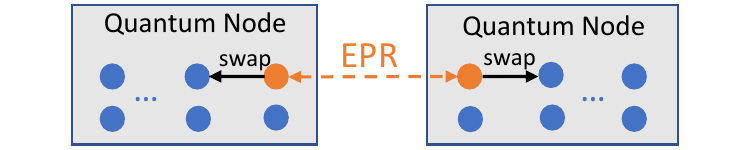}
\vspace*{-1\baselineskip}
  \caption{The concept of EPR pair buffering in a distributed quantum computing system.}
  \label{fig:comm_buffer}
\end{figure}

Efficient utilization of communication resources in quantum nodes requires a strategy for handling EPR pairs after generation. The concept of EPR pair buffering, proposed in~\cite{wu2022collcomm}, addresses this need by introducing a buffering mechanism to store generated EPR pairs. 
As illustrated in Fig.~\ref{fig:comm_buffer}, once an EPR pair is generated on the communication qubits (marked with orange color) of two connected quantum nodes, the entangled state is swapped into other available data qubits (marked with blue color) in the quantum node. This process releases communication qubits, enabling them to generate new EPR pairs. The buffered EPR pairs remain available for future quantum communication tasks.

\begin{figure}[h]
    \centering
    \includegraphics[width=\linewidth]{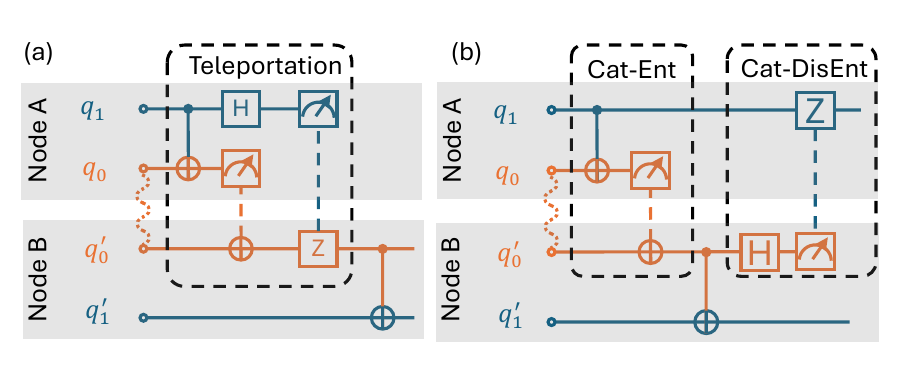}
\vspace*{-1\baselineskip}
    \caption{Two communication protocols. (a) TP-Comm version for implementing a remote CNOT. (b) CAT-Comm version for implementing a remote CNOT.}
    \label{fig:tp-cat-comm}
\end{figure}

\subsection{Quantum Communication Protocols}\label{subsec:comm_protocol}
With shared EPR pairs between quantum nodes, two quantum communication protocols, TP-Comm and Cat-Comm, can be utilized for quantum communication and remote gate operations~\cite{haner2021QMPI, wu2022collcomm, wu2022autocomm, Wu2023QuComm}.

Fig.~\ref{fig:tp-cat-comm} shows the quantum circuits for these protocols. The TP-Comm protocol, shown in Fig.~\ref{fig:tp-cat-comm}(a), employs a prepared EPR pair to teleport a qubit state from node A to node B. Once the state is teleported, the qubit in node B can serve as the control qubit in a CNOT gate, effectively realizing a remote CNOT gate between the two nodes.

In comparison, the Cat-Comm protocol,  shown in Fig.\ref{fig:tp-cat-comm}(b), utilizes a shared EPR pair to entangle the qubit in node B with the data qubit in node A through the ``cat-entangler'' circuit. After generating this entanglement, the qubit $q_0'$ in node B acts as an intermediary for the remote CNOT operation. Subsequently, the ``cat-disentangler'' circuit disentangles the qubit $q_0'$ in node B from the other two qubits to complete the remote CNOT operation. Unlike the TP-Comm protocol, the Cat-Comm protocol does not teleport the quantum state, which restricts the supported remote gate operations. However, Cat-Comm enables ``broadcasting'' of entanglement, making it suitable for multi-node or collective operations\cite{haner2021QMPI, wu2022collcomm}.

\subsection{Related Work}
\textbf{Quantum Networks: }
Quantum networks can be classified into two types. The first type, \emph{quantum internet}~\cite{Pirandola2016internet,Stephanie2018internet}, enables long-range quantum communication over optical links. These networks rely on quantum repeaters to mitigate photon loss. The second type, \emph{quantum intranet}~\cite{chow2021DQC}, enables quantum information transfer within shorter-range networks. Experimental work~\cite{Magnard2020microwave, Storz2023} has demonstrated that superconducting qubits connected via microwave links can realize such short-range quantum networks. Our work focus on quantum supercomputers where each node in the quantum supercomputer is connected via short range microwave links.

\textbf{DQC Architectures: }
Several scalable architectural directions have been explored to realize DQC. The \emph{chiplet architecture}~\cite{Smith2022Chiplet, LaRacuente_2025} integrates multiple smaller quantum chips via microwave links, analogous to multi-core classical systems where multiple processing cores collaborate to solve a problem in a distributed fashion. Another scalable design is the \emph{quantum data center (QDC)}~\cite{shapourian2025quantumdatacenterinfrastructures}, which utilizes optical switches to interconnect multiple quantum processors. QDCs operate over short distances, therefore avoid the need for quantum repeaters, making them well suited with near-term hardware capabilities~\cite{zhang2024QDC}.
The distinction between these approaches lies in their scaling philosophy: chiplet architectures achieve performance improvements by integrating multiple quantum processing elements within a single node (scale-up), while QDC architectures enhance computational capacity by interconnecting multiple independent quantum nodes (scale-out). Our work follows the scale-out paradigm, focusing specifically on accelerating quantum communication between nodes through the design of specialized hardware, the Quantum Network Processor Unit (QNPU).

\textbf{DQC Compilers: } 
A variety of compiler frameworks~\cite{wu2022autocomm, wu2022collcomm, Wu2023QuComm} have been proposed to optimize inter-node communications in distributed quantum systems. Other works focus on specific architecture setting: Zhang et al.\cite{zhang2024QDC} presents a compiler framework for EPR scheduling to mitigate entanglement generation latency in quantum data centers, and Zhang et al.~\cite{Zhang_2024_MECH} presents Multi-Entry Communication Highway (MECH), which leverages compiler-managed ancillary qubits to enhance concurrency in chiplet architectures. 
While these efforts improve communication efficiency through software-level optimizations, our work introduces hardware acceleration via the Quantum Network Processing Unit (QNPU), a design that both complements and benefits from compiler-driven strategies.

\textbf{Instruction Set Architecture and Microarchitecture for Quantum Communications: }
QuMIS~\cite{Fu2017QuMA} and eQASM~\cite{Fu2019eQASM} are executable ISAs designed for quantum processors, focusing on local quantum operations but lacking support for quantum communication between distributed processors.
NetQASM~\cite{Dahlberg2022NetQASM} extends this by supporting quantum network applications with instructions for generating entanglement between remote quantum nodes. 
More recently, Donne et al.~\cite{Donne2025OS} introduced an architecture and operating system (QNodeOS) to support quantum network applications with multitasking capabilities. Both the NetQASM and QNodeOS introduce the concept of Quantum Network Processing Unit (QNPU) for executing applications on quantum network nodes, and their QNPUs are a unified processing unit that processes both local quantum operations and remote quantum operations.
In contrast, our work takes a fundamentally different approach by physically decoupling computation and communication processing units, where QPUs are responsible for local computation and QNPUs are responsible for quantum communication operations.

\textbf{Large-Scale Fault-Tolerant Distributed Quantum Computer: } Kim et al.~\cite{Kim2024MQDQC} address the challenges of fault tolerance in million-qubit-scale quantum computers distributed across multiple dilution refrigerators. Their work focuses on implementing fault-tolerant quantum computation using surface codes with lattice surgery, with an emphasis on minimizing logical error rates in large-scale distributed systems. In comparison, our approach tackles the complementary challenge of providing efficient quantum communication between quantum nodes through a dedicated Quantum Network Processing Unit (QNPU). Their proposed fault-tolerant techniques could be integrated with our QNPU-based communication stack to construct a scalable distributed quantum system that is both fault-tolerant and communication-efficient.

\section{Quantum Supercomputer with Decoupled Processing Units}
In this section, we present the layered stack for the quantum supercomputer, then focus on the decoupled processing units of the quantum supercomputer.
\subsection{Layer Stack for Quantum Supercomputers}\label{subsec:layer_arch}
\begin{figure}[h]
    \centering
    \includegraphics[width=0.8\linewidth]{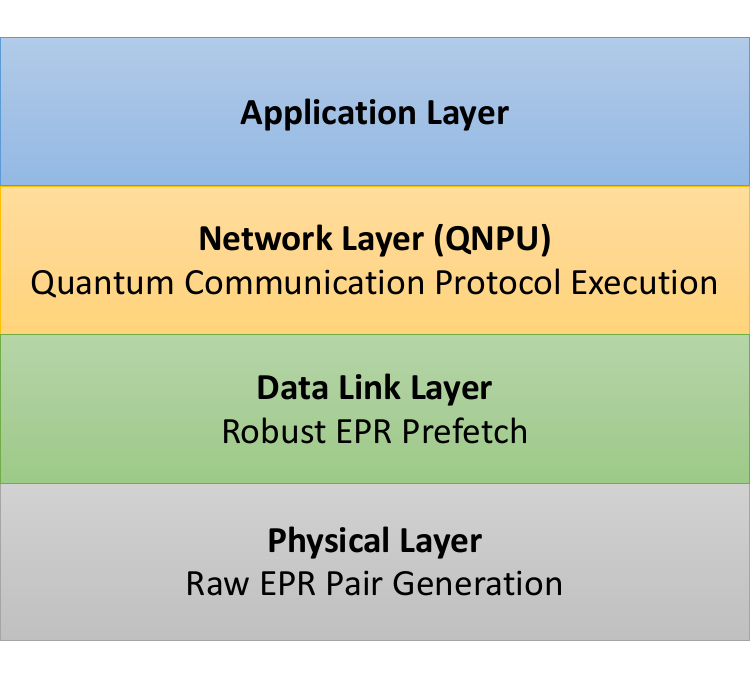}
\vspace*{-1\baselineskip}
    \caption{Layered architecture for quantum supercomputers. The design consists of four distinct layers, each providing specific functionalities and interfacing with adjacent layers.}
    \label{fig:network}
\end{figure}

Inspired by the classical network architecture’s Open Systems Interconnection (OSI) model, previous works have proposed layered quantum network stacks to support quantum network applications~\cite{Dahlberg2019linklayer_protocol,Kozlowski_2020}. Similarly, the work in~\cite{ang2024ARQUIN} introduced a layer stack for a multinode quantum computer (MNQC) architecture. Our proposed layer stack for quantum supercomputers draws upon these ideas while introduces dedicated hardware acceleration for communication protocols (through the QNPU at the network layer), proactive EPR resource management (EPR prefetch at the data link layer), and a clear architectural separation between local quantum computation and inter-node communication to enable efficient execution of distributed 
quantum algorithms.
Fig.~\ref{fig:network} illustrates our four-layer architecture.

\textbf{Physical Layer:} The physical layer forms the foundation of the quantum supercomputer. It is responsible for generating raw EPR pairs between directly connected quantum nodes. As our design adopts the established protocols~\cite{Dahlberg2019linklayer_protocol,Kozlowski_2020} for the physical layer, we do not further elaborate on its specifics.

\textbf{Data Link Layer:} We define the data link layer's primary function as EPR prefetch, which ensures reliable pre-generation and buffering~\cite{wu2022collcomm} of required EPR resources to hide EPR pair generation latency. Prior work~\cite{zhang2024QDC} has proposed compiler frameworks enables look-ahead scheduling of EPR pairs, which can be integrated into the data link layer to realize this prefetch functionality.

To efficiently manage and track prefetched EPR pairs, we introduce the EPR Resource Table, as shown in Table~\ref{tab:EPR_info_table}. Each quantum node maintains this table to track EPR resources. When an EPR pair is prefetched, corresponding entries are appended to the tables in both the source and destination nodes. Each entry contains a unique EPR Pair ID, the remote node identifier, the current state of the EPR pair (available, occupied, or empty), and the EPR qubit index indicating the location of the entangled qubit in the local node.

\begin{table}[h!]
\centering
\caption{Structure of the EPR Resource Table, where each row represents an EPR pair entry. The Pair ID uniquely identifies each EPR pair, while the Remote Node field indicates the connected quantum node. The State field can be "Available" (ready for use), "occupied" (currently in use), or "Empty" (available for new EPR pair prefetch). EPR Qubit Index showing the location of the entangled qubit in the local node. Empty entries (marked with "-") indicate slots available for new EPR pair prefetching.
}
\vspace*{-1\baselineskip}

\label{tab:EPR_info_table}
\resizebox{0.48\textwidth}{!}{%
\begin{tabular}{|c|c|c|c|c|c|}
\hline
\textbf{Pair ID} & \textbf{Remote Node} & \textbf{State} & \textbf{EPR Qubit Index}\\ \hline
1                & Node B              & Available      & 3                   \\ \hline
2                & Node C              & Occupied       & 7                   \\ \hline
-                & -             & Empty          & -                  \\ \hline
\end{tabular}
}
\end{table}

\textbf{Network Layer:} Operating above the data link layer, the network layer is responsible for executing quantum communication protocols including the TP-Comm and Cat-Comm protocols by utilizing the prefetched EPR resources. Upon receiving a protocol execution request, this layer consults the EPR Resource Table to identify and allocate a suitable EPR pair. Given that distributed quantum algorithms often involve numerous remote gates (operations that span spatially separated nodes), the network layer plays a crucial role in facilitating these communication-intensive operations.

To handle the high volume of such communication protocols, we introduce a dedicated hardware accelerator: the Quantum Network Processor Unit (QNPU). By offloading communication tasks to the QNPU within the network layer, our decoupled architecture allows the Quantum Processing Unit (QPU) to focus exclusively on local computations.

\textbf{Application Layer:} At the top of the stack, the application layer hosts the QPU, which runs distributed quantum algorithms. When a remote operation is required during algorithm execution, the QPU issues a communication request to the network layer. The QNPU processes this request, executing the requisite quantum communication protocol using the available EPR resources. This separation ensures that local computations and network operations are handled independently, maximizing overall system efficiency.

Our layered approach provides a comprehensive framework for quantum supercomputers, balancing the requirements of quantum communication, EPR resource management, and algorithm execution in a scalable and efficient manner.

\subsection{Decoupled Processing Units for Quantum Supercomputers}\label{subsec:network_layer_qnpu}
\begin{figure}[h]
  \centering
  \includegraphics[width=\linewidth]{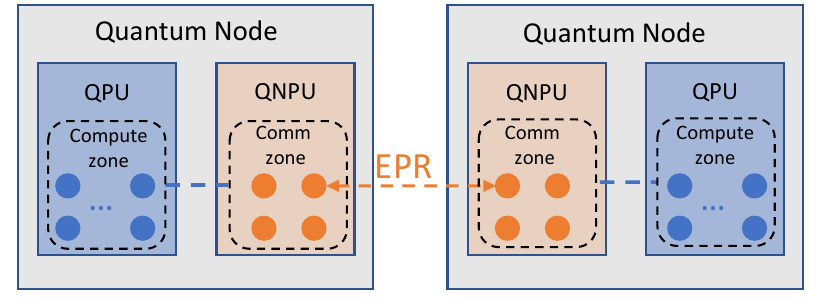}
\vspace*{-1\baselineskip}
  \caption{Decoupled processing units for quantum supercomputers. Each quantum node separates computation and communication resources: QPU contains a compute zone with data qubits for local quantum computation, while QNPU contains a communication zone with qubits for EPR pair prefetching and quantum communication protocol execution. The orange dashed line indicates EPR pair establishment between QNPUs, and blue dashed lines show qubit movement capabilities between the computation zone and communication zone.
  }
  \label{fig:new_epr_buffer}
\end{figure}
As established in Section~\ref{subsec:layer_arch}, our architecture employs a clear separation of concerns: the application layer hosts Quantum Processing Unit (QPU) for executing distributed quantum algorithms, when the QPU encounters operations requiring quantum communication, it offloads these tasks to the QNPU. This decoupled design contrasts with traditional designs \cite{Dahlberg2022NetQASM} that consolidate both computation and communication functions within a single processing unit.

In our design as shown in Fig.~\ref{fig:new_epr_buffer}, the QPU’s qubits constitute the "computation zone," dedicated solely to executing local quantum operations, whereas the QNPU’s qubits form the "communication zone," optimized for performing quantum communication protocols. 
This decoupled architecture offers significant advantages for both homogeneous and heterogeneous quantum computing scenarios. By separating computation from communication, QPU and QNPU can be specialized and optimized for their specific tasks. The superscalar QNPU pipeline design proposed in Section VI does not affect the QPU, as a result of this decoupled architecture.
Moreover, by decoupling the QPU from network communication responsibilities, different nodes in a quantum supercomputer could employ fundamentally different quantum technologies (such as superconducting circuits, trapped ions, or neutral atoms) for their local computations, while QNPUs provide a communication interface across the system. 
The clean separation of concerns not only simplifies current system design and implementation, but also provides a practical pathway toward realizing the heterogeneous quantum systems envisioned by the quantum computing community.

\section{Programming Model for Distributed Quantum Computing}
While distributed quantum computing offers significant advantages for scaling quantum applications, it introduces complexity in programming and EPR resource management. In this section, we introduce a programming model which extends OpenQASM~\cite{Cross2022QASM} to simplify the development of distributed quantum algorithms.

\subsection{DistQASM: Extending OpenQASM for Distributed Quantum Computing}
We propose DistQASM, an extension of OpenQASM~\cite{Cross2022QASM} designed specifically for distributed quantum computing. DistQASM builds upon OpenQASM's human-readable representation of quantum circuits while adding critical features for distributed quantum computing. By extending rather than replacing OpenQASM, DistQASM maintains compatibility with the existing quantum software ecosystem while enabling the expression of distributed quantum operations. We introduce several key extensions in DistQASM:
\begin{enumerate}
\item \textbf{Node-specific qubit declarations} to identify which physical node hosts each qubit;
\item \textbf{Explicit Remote Block Markers} to clearly delineate code that will execute on a remote node;
\item \textbf{Direct Protocol Specification} to define instructions for quantum communication protocols so that user can directly choose the protocol when writing the distributed quantum program. For the TP-Comm protocol, we define the \textbf{\texttt{teleport}} instruction to teleport a state from the source node to the destination node. For the CAT-COMM protocol, we define two DistQASM instructions: \textbf{\texttt{cat\_ent}} and \textbf{\texttt{cat\_disent}}. The cat\_ent instruction entangles the qubit in the source node with the qubit in the destination node. The cat\_disent disentangles these two qubits.
\end{enumerate}
The following are the examples of a distributed quantum program (GHZ) using DistQASM with two different protocols:
\begin{lstlisting}[caption={A GHZ state preparation circuit distributed in quantum node A and B using the TP-Comm protocol}, label=lst:DistQASM1]
qreg q[2] @nodeA;//Declare two qubits located on Node A
qreg r[2] @nodeB;//Declare two qubits located on Node B
h q[0];//local operations on Node A            
cnot q[0], q[1];//local operations on Node A     
teleport q[1], r[0];//Use TP-Comm protocol to transfer state from Node A to B 
pragma remote_begin nodeB //Indicates that subsequent operations execute on Node B
  cnot r[0], r[1];//local operation on Node B    
pragma remote_end //Return execution control to Node A
\end{lstlisting}
\begin{lstlisting}[caption={A GHZ state preparation circuit distributed in quantum node A and B using the CAT-Comm protocol}, label=lst:DistQASM2]
qreg q[2] @nodeA;//Declare two qubits located on Node A
qreg r[2] @nodeB;//Declare two qubits located on Node B
h q[0];//local operations on Node A                    
cnot q[0], q[1];//local operations on Node A          
cat_ent q[1], r[0];//CAT-Entanglement to establish remote entanglement  
pragma remote_begin nodeB //Indicates that subsequent operations execute on Node B
  cnot r[0], r[1];//local operation on Node B   
pragma remote_end //Return execution control to Node A
cat_disent r[0], q[1];//CAT-Disentanglement to do state disentanglement
\end{lstlisting}

\subsection{DistQASM Instruction Translation}\label{subsec:distQASMtrans}
As shown in Listings~\ref{lst:DistQASM1} and~\ref{lst:DistQASM2}, DistQASM supports user-friendly instructions for both TP-Comm and CAT-Comm protocols. These high-level instructions abstract away the underlying EPR resource management and communication protocol coordination between quantum nodes.

In our design, the compiler translates these \textbf{high-level} DistQASM instructions into \textbf{low-level} instruction sequences executed on respective nodes' QPUs. The \texttt{teleport} instruction is translated into two complementary instruction sequences: one instruction sequence on the source node and the other on the destination node. Similarly, the \texttt{cat\_ent} or the \texttt{cat\_disent} is also translated into paired instruction sequences that coordinate across the participating nodes. We illustrate this translation process using the \texttt{teleport} instruction as an example:

On the source node's QPU, the compiler generates an instruction sequence that performs:
\begin{enumerate}
    \item \textbf{Qubit allocation and zone transition}: An available qubit is requested from the communication zone, and the qubit state is transitioned from the computation zone to the communication zone. This step reflects our decoupled QPU architecture, where computation and communication resources are physically separated. We assume that the QPU ISA provides instructions (e.g., SWAP or physical move) to perform this zone transition.
    \item \textbf{Teleportation protocol execution}: The teleportation protocol is initiated and delegated to the QNPU. We define an ISA instruction, \texttt{SEND\_TP\_QUBIT}, to represent this step. Details are provided in the next section of QNPU ISA design.
\end{enumerate}

On the destination node's QPU, a complementary instruction sequence performs:
\begin{enumerate}
    \item \textbf{Reception of the teleported state}: The state is received via the \texttt{GET\_TP\_QUBIT} instruction, which is executed by the QNPU. Details about this instruction's definition are provided in the next section of QNPU ISA design.
    \item \textbf{Zone transition back to computation zone}: Once finished the teleportation, the QNPU signals the QPU via a TRANSFER\_SUCCESS\_NOTIFY instruction (Details of this instruction are provided in Section~\ref{subsec:micro-ops}). The QPU then initiates a zone transition to move the teleported qubit state from the QNPU’s communication zone to the QPU’s computation zone. 
\end{enumerate}

This instruction translation strategy allows DistQASM to serve as a scalable and expressive programming interface for distributed quantum systems, while delegating low-level resource management and synchronization to the compiler and underlying microarchitecture.

\section{QNPU ISA Design}
The QNPU Instruction Set Architecture (ISA) is designed to efficiently execute the quantum communication protocols, TP-Comm or Cat-Comm. 
Our QNPU ISA design is guided by two principles:
\begin{enumerate}
    \item \textbf{Abstraction of Quantum Communication Protocols:} 
    Instructions related to communication protocols initially flow from the QPU to the QNPU for execution, and QPU does not need to know the implementation details of the communication protocol. Therefore, our ISA should provides concise, well-defined protocol instructions that enable efficient instruction transfer between these components while abstracting the underlying complexity of quantum communication.  
    \item \textbf{Protocol Implementation with $\mu$ops:} When the QNPU receives the communication protocol instructions from the QPU, it must execute the corresponding sequence of operations required for the protocol implementation. To achieve this, our architecture adopts the microcode approach, allowing the QNPU to decode the communication protocol instructions into sequences of micro-operations ($\mu$ops) for protocol execution. Therefore, the ISA must define these $\mu$ops, including EPR pair management, quantum gate operations, measurements, and classical message exchange between nodes. 
\end{enumerate}

\subsection{Communication Protocol Instructions}
We define the instructions for both the TP-Comm and Cat-Comm protocols. These instructions are used by the compiler to support the DistQASM communication instructions, as discussed in Section~\ref{subsec:distQASMtrans}. When encountered by the QPU, these instructions are sent to QNPU for execution.
\begin{itemize}
    \item \texttt{SEND\_TP\_QUBIT <qubit\_reg>, <dst\_node\_id>}: Executes the TP-Comm protocol to transfer a quantum state from a source qubit in the source node to the destination node. This instruction handles the operations in the source node.
    \item \texttt{GET\_TP\_QUBIT <qubit\_reg>}: This instruction is a complementary instruction to \texttt{SEND\_TP}, which handles the operations in the destination node. It identifies the teleported qubit and saves its index into the specified register, enabling the destination QPU to access the teleported state register.
    \item \texttt{SEND\_CAT\_ENT\_QUBIT <qubit\_reg>, <dst\_node\_id>}: Executes the CAT-Comm protocol to entangle a quantum state from a qubit in the source node to the destination node. This instruction handles the operations in the source node.
    \item \texttt{GET\_CAT\_ENT\_QUBIT <qubit\_reg>}:  This instruction is a complementary instruction to \texttt{SEND\_CAT\_ENT}, which handles the operations in the destination node. It identifies the entangled qubit and saves its index into the specified register for subsequent operations by the destination QPU.
    \item \texttt{SEND\_CAT\_DISENT\_QUBIT <qubit\_reg>, <src\_node\_id>}: Executes the disentanglement part of the CAT-Comm protocol to dis-entangle the qubit from the destination node to the source node.
    \item \texttt{GET\_CAT\_DISENT\_QUBIT <qubit\_reg>}:  This instruction is a complementary instruction to \texttt{SEND\_CAT\_DISENT}.
\end{itemize}

\subsection{Micro-Operations }\label{subsec:micro-ops}
When the QNPU receives the communication protocol instructions from the QPU, it decodes them into sequences of micro-operations ($\mu$ops) to implement the protocol. Here, we define the comprehensive set of $\mu$ops required for quantum communication protocols and divide them into three categories: EPR Resource Management $\mu$ops, Classical Communication $\mu$ops, and Quantum $\mu$ops.

\subsubsection{EPR Resource Management $\mu$ops}
These $\mu$ops manage the entanglement resources necessary for distributed quantum communication:
\begin{itemize}
    \item \texttt{EPR\_RESERVE <EPRIdReg>, <dst\_node\_id>}: 
    The source node's QNPU executes this $\mu$op to occupy a prefetched EPR pair for quantum communication by checking the EPR resource table. If an available EPR pair is found, its ID is saved into the register EPRIdReg.
    \item \texttt{EPR\_RESERVE\_SYNC <StatusReg>}: 
    The destination node's QNPU executes this $\mu$op to synchronize EPR information with the source node, with the synchronization status saved in StatusReg. This ensures resource consistency across the distributed quantum nodes.
    \item \texttt{GET\_EPR\_QUBIT <qubit\_reg>}:
    The QNPU executes this $\mu$op to retrieve the local qubit associated with the occupied EPR pair, storing its address in qubit\_reg, so that later on the quantum operations can be applied to this qubit.
    \item \texttt{EPR\_RELEASE <qubit\_reg>}:
    When the qubit in qubit\_reg is measured or its state is transferred, the QNPU executes this $\mu$op to update the EPR resource table, marking the EPR pair's status as empty and releasing the qubit for future use. This supports efficient resource recycling.
\end{itemize}
\subsubsection{Classical Communication $\mu$ops}
These $\mu$ops manage the classical information exchange required to coordinate quantum operations across distributed nodes.

\subsubsection{Quantum $\mu$ops}
Quantum $\mu$ops implement the local quantum operations required for communication protocols. These include standard quantum gates and measurements as described in previous works~\cite{Fu2017QuMA,Fu2019eQASM}. With our execution mdoel, these $\mu$ops are executed by the QNPU specifically for communication-related quantum operations, while the QPU handles computation-zone quantum operations.

\textit{Example:}
To illustrate how the communication protocol instructions are decoded into $\mu$ops, we present the $\mu$op sequences for the TP-Comm protocol related instructions, which include the two complementary instructions,  \texttt{SEND\_TP\_QUBIT} and \texttt{GET\_TP\_QUBIT}. 

The source node's QNPU would execute the following sequence of $\mu$ops upon receiving the instruction \texttt{SEND\_TP\_QUBIT} from its QPU:

\begin{lstlisting}[caption={Micro-operation sequence for SEND\_TP\_QUBIT}, label=lst:tp_comm_uops]
// 1. EPR resource management
EPR_RESERVE  EPRIdReg, NodeB// Reserve an EPR pair betweem Node A and Node B for the TP-COMM protocol, store the EPR id into EPRIdReg
SEND_EPR_ID NodeB, TransferID_1, [EPRIdReg] // Classical communication, send EPR ID to Node B for synchronization
ACK_WAIT NodeB, TransferID_1, StatusReg//Wait for acknowledgment from node B to confirm successful EPR pair synchronization
GET_EPR_QUBIT EPRQubReg //get the EPR pair's local qubit, result stored in EPRQubReg
// 2. perform quantum operations and measurements
CNOT CommQubReg, EPRQubReg  // Apply CNOT gate
H CommQubReg                 // Apply Hadamard gate
MEAS EPRQubReg, BitXReg   // Measurement, result stored in BitXReg
MEAS CommQubReg, BitZReg      // Measurement, result stored in BitZReg
// 3. Release the local qubit for the current reserved EPR pair
EPR_RELEASE EPRQubReg //After measurement, EPR qubit can be released
// 4. Classical communication of measurement results
TP_SEND_BITS NodeB, TransferID_1, BitZ, BitX  // Send measurement results to Node B
\end{lstlisting}

In Listing 3, \texttt{EPR\_RESERVE}, \texttt{GET\_EPR\_QUBIT}, and \texttt{EPR\_RELEASE} are EPR management $\mu$ops; \texttt{SEND\_EPR\_ID}, \texttt{ACK\_WAIT}, and \texttt{TP\_SEND\_BITS} are classical communication $\mu$ops; and \texttt{CNOT}, \texttt{H}, \texttt{MEAS} are quantum $\mu$ops. They are colored coded in the listings.

Similarly, the destination node's QNPU would execute the following sequence of $\mu$ops upon receiving the instruction \texttt{GET\_TP\_QUBIT}:
\begin{lstlisting}[caption={Micro-operation sequence for GET\_TP\_QUBIT}, label=lst:tp_receive_uops]
// 1. EPR resource management
RECV_EPR_ID EPRIdReg, TransferID_1 // Receive EPR ID information, result stored in EPRIdReg
EPR_RESERVE_SYNC StatusReg //Verify EPR pair availability and update local EPR information table to be reserved
ACK_SEND NodeA, TransferID_1, StatusReg //Send acknowledgment to Node A regarding EPR pair synchronization status
GET_EPR_QUBIT TeleportQubReg //Get local EPR qubit for teleportation, result stored in TeleportQubReg
// 2. Apply quantum operations based on classical information
TP_RECV_BITS BitXReg, BitZReg TransferID_1  // Receive measurment information from Node A, results stored in the BitXReg, and BitZReg
(BitXReg) X TeleportQubReg  // Apply X gate if BitX is 1
(BitZReg) Z TeleportQubReg   // Apply Z gate if BitZ is 1
// 3. Update teleportation tracking information
TRANSFER_SUCCESS_NOTIFY TransferID_1   // Notify Node B's QPU of teleportation success, enables zone transition to computation zone
// 4. Release the qubit for the EPR pair(after zone transition completes)
EPR_RELEASE TeleportQubReg //Release qubit after state transfer to computation zone
\end{lstlisting}

The micro-operation sequences for the CAT-Comm protocol would follow a similar pattern.
This detailed design of $\mu$ops provides the QNPU with a comprehensive set of low-level operations to implement quantum communication protocols, which allows for flexibility in protocol implementation while maintaining a clean interface for QPU.

\section{QNPU Microarchitecture}
\subsection{QNPU Scalar Pipeline Design}
\begin{figure*}
  \centering
  {
  \includegraphics[width=\linewidth]{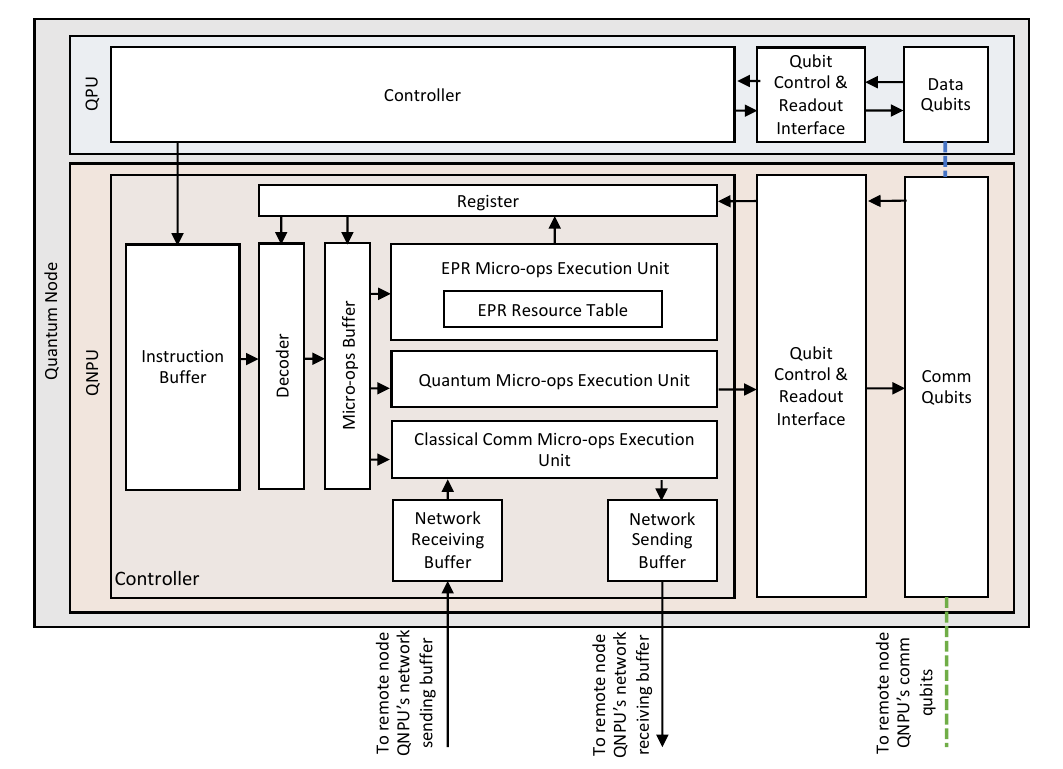}
  }\hfil
  \vspace*{-1\baselineskip}
  \caption{Overview of the QNPU microarchitecture. The primary QPU's controller orchestrates the execution across nodes, while the QNPU handles communication protocols through specialized execution units for EPR resource management, quantum operations, and classical communication.}
  \label{fig:arch}
\end{figure*}
\begin{figure*}
  \centering
  {
  \includegraphics[width=\linewidth]{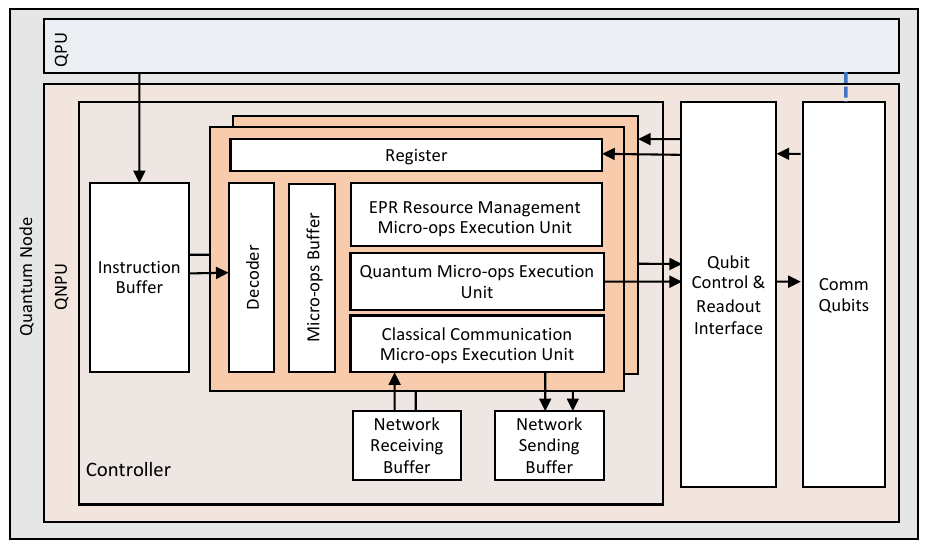}
  }\hfil
  \vspace*{-1\baselineskip}
  \caption{Overview of the QNPU Superscalar microarchitecture. This figure shows a two-way superscalar design which enables parallel execution of two communication protocol instructions. }
  \label{fig:arch_superscalar}
\end{figure*}

As illustrated in Fig.~\ref{fig:arch}, each quantum node in our proposed architecture consists of both a Quantum Processing Unit (QPU) and a Quantum Network Processing Unit (QNPU).

The QPU comprises three main components: (1) a classical controller, (2) a qubit control and readout interface for the data qubits in the computation zone, and (3) data qubits in the computation zone. The classical controller processes input instructions and translates quantum operations into codewords, which are then sent to the qubit control and readout interface. This interface is responsible for converting the received codewords into control pulses for the qubits in the computation zone. Additionally, it performs qubit measurements and sends the results back to the classical controller.

The QNPU serves as the communication interface between quantum nodes, enabling the execution of quantum communication protocols. Similar to the QPU, the QNPU consists of three components: (1) a classical controller, (2) a qubit control and readout interface for the qubits in the communication zone and (3) qubits in the communication zone. The QNPU classical controller receives the quantum communication protocol instructions from its QPU and decodes them into sequences of $\mu$ops. These $\mu$ops are then executed by specialized execution units and ultimately translated into control pulses for the qubits in the communication zone.

While previous works~\cite{Fu2017QuMA, Fu2019eQASM, Zhang2021parallelArch} have discussed QPU microarchitecture design, this paper presents a novel QNPU microarchitecture specifically designed for distributed quantum computing. Our proposed pipeline architecture, shown in Fig.~\ref{fig:arch}, efficiently processes the communication protocol instructions. 
The key components of our QNPU microarchitecture include:
\begin{itemize}
\item \textbf{Instruction Buffer:} Stores quantum communication instructions delegated from the QPU. This buffer enables instructions from the QPU to its QNPU.
\item \textbf{Decoder:} Translates the communication protocol instructions into sequences of $\mu$ops that are placed in the $\mu$ops buffer. The decoder maintains instruction atomicity by waiting for all $\mu$ops from the current instruction to complete execution before decoding a new instruction.
\item \textbf{$\mu$ops Buffer:} Holds the decoded $\mu$ops before they are dispatched to their respective execution units. This buffer helps manage the flow of operations across the specialized execution units.
\item \textbf{Specialized Execution Units:} Three distinct execution units directly correspond to the three categories of $\mu$ops defined in our ISA:
\begin{itemize}
    \item \textit{EPR Resource Management $\mu$ops Execution Unit:} Manages the entanglement resources through the EPR resource table, which tracks the status and allocation of EPR pairs.
    \item \textit{Classical Communication $\mu$ops Execution Unit:} Handles the exchange of classical information between nodes, coordinating with the network buffers.
    \item \textit{Quantum $\mu$ops Execution Unit:} Executes quantum operations on qubits in the communication zone.
\end{itemize}
\item \textbf{Register File:} Provides storage for operation results and state information, accessible by all execution units.
\item \textbf{Network Buffers:} The sending and receiving buffers enable classical message exchange between QNPUs across different nodes, ensuring proper coordination during protocol execution.
\end{itemize}

\paragraph{QNPU Micro-Operation Execution}
In order to show how the QNPU instructions run on the QNPU, we now present how the QNPUs for a source Node A and a destination Node B execute the complementary instructions, \texttt{SEND\_TP\_QUBIT} and \texttt{GET\_TP\_QUBIT}, respectively. The instructions for the Cat-Comm protocol execute in a similar way.

Node A's QNPU (\texttt{SEND\_TP\_QUBIT} Execution):
When Node A's QNPU receives the \texttt{SEND\_TP\_QUBIT} instruction, its decoder generates the $\mu$ops sequence as shown in Listing~\ref{lst:tp_comm_uops} and routes each $\mu$op to the appropriate execution unit:
\begin{enumerate}
\item The \texttt{EPR\_RESERVE} $\mu$op executes in the EPR Resource Management Execution Unit, allocating an EPR pair from the EPR Information Table and storing its ID in the register EPRIdReg with the ready bit set.
\item The \texttt{SEND\_EPR\_ID} $\mu$op runs in the Classical Communication Execution Unit, forwarding the EPR pair ID to Node B through the network sending buffer.
\item The \texttt{ACK\_WAIT} $\mu$op keep querying the network receiving buffer for synchronization status from Node B, storing the result in the StatusReg.
\item Upon successful synchronization, \texttt{GET\_EPR\_QUBIT} retrieves the local qubit for the EPR pair by consulting the EPR resource table.
\item The Quantum $\mu$ops Execution Unit processes the CNOT, Hadamard, and measurement operations sequentially.
\item After measurements, \texttt{EPR\_RELEASE} updates the EPR resource table to mark the qubit as available for future use.
\item Finally, \texttt{TP\_SEND\_BITS} transmits the measurement results to Node B, completing the portion of the teleportation protocol in the source node.
\end{enumerate}

Node B's QNPU (\texttt{GET\_TP\_QUBIT} Execution):
Concurrently, Node B's QNPU processes the complementary \texttt{GET\_TP\_QUBIT} instruction:
\begin{enumerate}
\item The \texttt{RECV\_EPR\_ID} $\mu$op waits for EPR ID information from Node A and stores it in EPRIdReg with the ready bit set.
\item The \texttt{EPR\_RESERVE\_SYNC} $\mu$op verifies EPR pair availability and updates the local EPR resource table to match Node A's record.
\item \texttt{ACK\_SEND} confirms successful synchronization back to Node A, allowing the protocol to proceed.
\item \texttt{GET\_EPR\_QUBIT} identifies the local EPR qubit for the teleportation operation.
\item \texttt{TP\_RECV\_BITS} receives the measurement results from Node A, enabling the conditional application of X and Z gates based on these results.
\item Upon completion of the quantum operations, \\\texttt{TP\_SUCCESS\_NOTIFY} informs Node B's QPU that the teleportation has succeeded, enabling zone transition.
\item Finally, after the state transfer is complete, \texttt{EPR\_RELEASE} frees the EPR resource for future use.
\end{enumerate}

\subsection{QNPU Superscalar Design}
Distributed quantum circuits frequently require remote gate operations, resulting in numerous quantum communication protocol instructions. Our scalar design, while effective, creates a performance bottleneck as the QNPU decoder must wait for all $\mu$ops from the current instruction to complete execution before decoding a new coomunication protocol instruction. This sequential processing limits throughput, especially in communication-intensive quantum algorithms. 

To address this limitation, we propose a superscalar pipeline design for the QNPU that enables parallel execution of multiple communication protocol instructions. 
Note that as the communication protocol instructions are first processed in QPU, QPU checks the data dependencies the same way as classical instructions and only issues the ready instructions. As a result, only independent communication protocol instructions are forwarded to the QNPU for concurrent execution, ensuring correctness while maximizing parallelism.

As shown in Fig.~\ref{fig:arch_superscalar}, our superscalar QNPU architecture incorporates multiple Decoder where each decoder independently processes communication protocol instructions, allowing simultaneous decoding of multiple instructions. Each decoder has its own $\mu$ops buffer and execution units, preventing resource contention between parallel instruction streams.
This architecture offers significant advantages for distributed quantum computing. When one decoder stalls due to outstanding $\mu$ops in its buffer, incoming communication protocol instructions can be routed to available decoders. 

This superscalar design is particularly useful for executing distributed quantum circuits containing multiple independent remote quantum operations. By processing multiple communication protocol instructions in parallel, the superscalar QNPU significantly reduces the overall execution time, accelerating distributed quantum algorithms.

\section{Evaluation}\label{sec:eva}
\subsection{Experiment Setup}
\textbf{Performance Simulator: }  
We developed a cycle-level simulator to evaluate the timing behavior of our proposed decoupled architecture. The simulator can be configured to model both the scalar and superscalar QNPU designs. We assume the availability of a compiler that efficiently schedules EPR pair generation, as demonstrated in prior work~\cite{zhang2024QDC}. With such compiler support, we model perfect EPR scheduling and pre-generation, which hides the latency of EPR generation.

\textbf{Baseline: }
For comparison, we implement a baseline architecture following the unified processing model of NetQASM~\cite{Dahlberg2022NetQASM}, where a single QNPU handles both local and remote quantum operations. 
Our baseline uses an in-order pipelined scalar QNPU that executes instructions for both local and remote gates. 
In other words, it supports pipelined execution but not parallel execution of instructions. For fairness, we assume this monolithic QNPU also has the same compiler support as our decoupled architecture, i.e., perfect EPR scheduling and pre-generation.

\textbf{Benchmarks: }
Our benchmarks include Hamiltonian Simulation, Greenberger–Horne–Zeilinger (GHZ), Bernstein-Vazirani (BV), Quantum Fourier Transform (QFT), Variational Quantum Eigensolver (VQE), and Quantum Approximate Optimization Algorithm (QAOA). For Hamiltonian simulation, we study one-dimensional Transverse Field Ising Models (TFIM) of varying sizes. For VQE, we adopt the two-local ansatz from the Qiskit circuit library, with full entanglement layers or linear entanglement layers. For QAOA, we evaluate MaxCut problems on 2-regular graphs and randomly generated graphs. For randomly generated graphs, half of all possible edges are connected. 

\textbf{Metric:}
We evaluate performance using \textit{Execution Time}, defined as the total time required for a distributed quantum program to complete on the given architecture.

\begin{table*}[htbp]
\centering

\caption{Performance Evaluation Results}
\vspace*{-1\baselineskip}
\label{tab:qnpu_results}
\resizebox{\textwidth}{!}{%
\begin{tabular}{|l|l|c|c|c|c|c|c|c|}
\hline
\textbf{Experiments} & \textbf{Benchmark} & \textbf{\#Tot.} & \textbf{\#Max.} & \textbf{Monolithic arch.:} & \textbf{Decoupled arch.} & \textbf{Imprv.} & \textbf{Decoupled arch.} & \textbf{Imprv.} \\
&\textbf{name-\#qubit-\#nodes} & \textbf{remote} & \textbf{remote} & \textbf{execution cycles} & \textbf{(scalar QNPU):} & & \textbf{(4-way}  &  \\
& & \textbf{CNOTs} & \textbf{CNOTs/node} & & \textbf{execution cycles} & & \textbf{superscalar QNPU):} & \\
&&&&&&&\textbf{execution cycles} &\\
\hline
\multirow{12}{*}{\begin{tabular}[c]{@{}l@{}}\textbf{Circuit-size Scaling} \\ Variable qubits, constant nodes\end{tabular}} 
& Hamiltonian Simulation-50-5 & 8 & 4 & 468 & 464 & 0.85\% & 464 & 0.85\% \\
\cline{2-9}
& Hamiltonian Simulation-100-5 & 8 & 4 & 768 & 764 & 0.52\% & 764 & 0.52\% \\
\cline{2-9}
& Hamiltonian Simulation-150-5 & 8 & 4 & 1068 & 1064 & 0.37\% & 1064 & 0.37\% \\
\cline{2-9}
& GHZ-50-5 & 4 & 2 & 188 & 184 & 2.13\% & 184 & 2.13\% \\
\cline{2-9}
& GHZ-100-5 & 4 & 2 & 288 & 284 & 1.39\% & 284 & 1.39\% \\
\cline{2-9}
& GHZ-150-5 & 4 & 2 & 388 & 384 & 1.03\% & 384 & 1.03\% \\
\cline{2-9}
& BV-50-5 & 40 & 40 & 546 & 528 & 3.30\% & 144 & 73.63\% \\
\cline{2-9}
& BV-100-5 & 80 & 80 & 1072 & 1051 & 1.96\% & 271 & 74.72\% \\
\cline{2-9}
& BV-150-5 & 120 & 120 & 1615 & 1573 & 2.60\% & 410 & 74.61\% \\
\cline{2-9}
& QFT-50-5 & 1020 & 408 & 24214 & 24006 & 0.86\% & 6846 & 71.73\% \\
\cline{2-9}
& QFT-100-5 & 4040 & 1616 & 94154 & 93248 & 0.96\% & 25520 & 72.90\% \\
\cline{2-9}
& QFT-150-5 & 9060 & 3624 & 210206 & 208220 & 0.94\% & 54884 & 73.89\% \\
\cline{2-9}
& VQE-linear-entanglement-50-5 & 4 & 2 & 188 & 184 & 2.13\% & 184 & 2.13\% \\
\cline{2-9}
& VQE-linear-entanglement-100-5 & 4 & 2 & 288 & 284 & 1.39\% & 284 & 1.39\% \\
\cline{2-9}
& VQE-linear-entanglement-150-5 & 4 & 2 & 388 & 384 & 1.03\% & 384 & 1.03\% \\
\cline{2-9}
& VQE-full-entanglement-50-5 & 1000 & 400 & 11866 & 11799 & 0.56\% & 3371 & 71.59\% \\
\cline{2-9}
& VQE-full-entanglement-100-5 & 4000 & 1600 & 46681 & 46220 & 0.99\% & 12680 & 72.84\% \\
\cline{2-9}
& VQE-full-entanglement-150-5 & 9000 & 3600 & 104489 & 103506 & 0.94\% & 27320 & 73.85\% \\
\cline{2-9}
& QAOA-50-5 & 2000 & 800 & 24383 & 23583 & 3.28\% & 6365 & 73.90\% \\
\cline{2-9}
& QAOA-100-5 & 8000 & 3200 & 97006 & 93354 & 3.76\% & 24372 & 74.88\% \\
\cline{2-9}
& QAOA-150-5 & 18000 & 7200 & 217525 & 209232 & 3.81\% & 54042 & 75.16\% \\
\hline
\multirow{12}{*}{\begin{tabular}[c]{@{}l@{}}\textbf{Node-count Scaling} \\ Variable nodes, constant qubits\end{tabular}} 
& Hamiltonian Simulation-150-2 & 2 & 2 & 945 & 944 & 0.11\% & 944 & 0.11\% \\
\cline{2-9}
& Hamiltonian Simulation-150-5 & 8 & 4 & 1068 & 1064 & 0.37\% & 1064 & 0.37\% \\
\cline{2-9}
& Hamiltonian Simulation-150-10 & 18 & 4 & 1273 & 1264 & 0.71\% & 1264 & 0.71\% \\
\cline{2-9}
& GHZ-150-2 & 1 & 1 & 325 & 324 & 0.31\% & 324 & 0.31\% \\
\cline{2-9}
& GHZ-150-5 & 4 & 2 & 388 & 384 & 1.03\% & 384 & 1.03\% \\
\cline{2-9}
& GHZ-150-10 & 9 & 2 & 493 & 484 & 1.83\% & 484 & 1.83\% \\
\cline{2-9}
& BV-150-2 & 75 & 75 & 1791 & 1675 & 6.48\% & 444 & 75.21\% \\
\cline{2-9}
& BV-150-5 & 120 & 120 & 1615 & 1573 & 2.60\% & 410 & 74.61\% \\
\cline{2-9}
& BV-150-10 & 135 & 135 & 1380 & 1371 & 0.65\% & 364 & 73.62\% \\
\cline{2-9}
& QFT-150-2 & 5700 & 5700 & 253238 & 252236 & 0.40\% & 65068 & 74.31\% \\
\cline{2-9}
& QFT-150-5 & 9060 & 3624 & 210206 & 208220 & 0.94\% & 54884 & 73.89\% \\
\cline{2-9}
& QFT-150-10 & 10200 & 2040 & 198044 & 196292 & 0.88\% & 46060 & 76.74\% \\
\cline{2-9}
& VQE-linear-entanglement-150-2 & 1 & 1 & 325 & 324 & 0.31\% & 324 & 0.31\% \\
\cline{2-9}
& VQE-linear-entanglement-150-5 & 4 & 2 & 388 & 384 & 1.03\% & 384 & 1.03\% \\
\cline{2-9}
& VQE-linear-entanglement-150-10 & 9 & 2 & 493 & 484 & 1.83\% & 484 & 1.83\% \\
\cline{2-9}
& VQE-full-entanglement-150-2 & 5625 & 5625 & 125103 & 124614 & 0.39\% & 32258 & 74.21\% \\
\cline{2-9}
& VQE-full-entanglement-150-5 & 9000 & 3600 & 104489 & 103506 & 0.94\% & 27320 & 73.85\% \\
\cline{2-9}
& VQE-full-entanglement-150-10 & 10125 & 2025 & 98679 & 97842 & 0.85\% & 22964 & 76.73\% \\
\cline{2-9}
& QAOA-150-2 & 11250 & 11250 & 251687 & 250135 & 0.62\% & 65707 & 73.89\% \\
\cline{2-9}
& QAOA-150-5 & 18000 & 7200 & 217525 & 209232 & 3.81\% & 54042 & 75.16\% \\
\cline{2-9}
& QAOA-150-10 & 20250 & 4050 & 208368 & 197034 & 5.44\% & 47602 & 77.15\% \\
\hline
\end{tabular}%
}
\end{table*}

\subsection{Experiment Results}\label{subsec:eva}
We evaluate the execution time of distributed quantum workloads under three architectural configurations: (1) a monolithic QNPU architecture (i.e., baseline), (2) a decoupled design with a scalar QNPU, and (3) a decoupled design with a 4-way superscalar QNPU. Table~\ref{tab:qnpu_results} summarizes the results of benchmarks with different circuit sizes and node counts.

\textbf{Decoupled vs. Monolithic Architectures: }
Across all benchmarks, the decoupled architecture with a scalar QNPU provides only marginal improvements over the monolithic baseline, typically around 1\%. This indicates that simply removing contention between computation and communication yields limited benefits, as the execution time of local gates is small compared to that of remote gates. In contrast, the decoupled architecture with a superscalar QNPU achieves substantial speedups, often exceeding 70\%, for communication intensive benchmarks like BV, QFT, VQE with full entanglement ansatz, and QAOA. These results show that the true advantage of our proposed decoupling architecture lies in separating computation from communication, which simplifies the design of superscalar communication pipelines and enables efficient parallel execution of remote gates.

\begin{figure*}[t]
    \centering
    \begin{subfigure}[t]{0.32\textwidth}
        \centering
        \includegraphics[width=\textwidth]{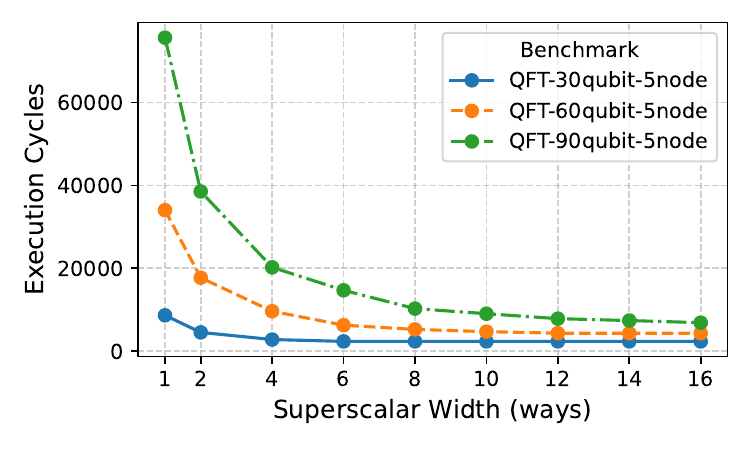}
        \vspace*{-1\baselineskip}
        \caption{QFT}
        \label{fig:qft_width}
    \end{subfigure}
    \begin{subfigure}[t]{0.32\textwidth}
        \centering
        \includegraphics[width=\textwidth]{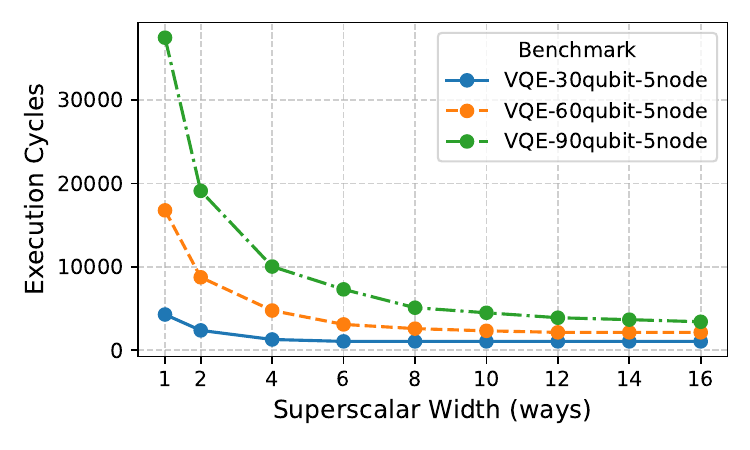}
        \vspace*{-1\baselineskip}
        \caption{VQE}
        \label{fig:vqe_width}
    \end{subfigure}
    \begin{subfigure}[t]{0.32\textwidth}
        \centering
        \includegraphics[width=\textwidth]{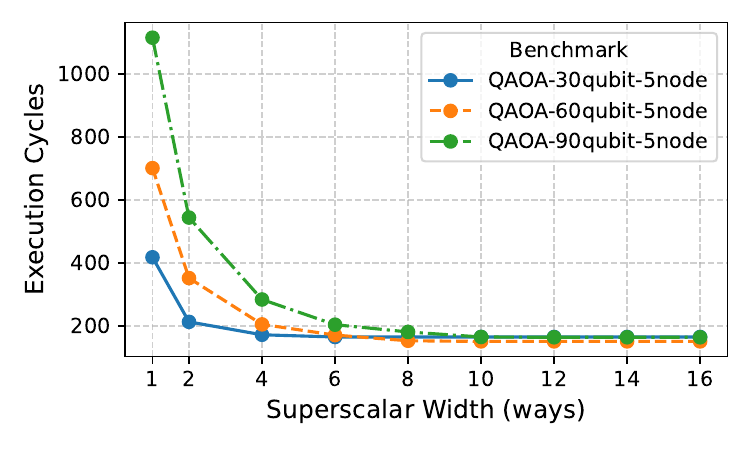}
        \vspace*{-1\baselineskip}
        \caption{QAOA}
        \label{fig:qaoa_width}
    \end{subfigure}
\vspace*{-1\baselineskip}    \caption{Impact of superscalar width on execution cycles across benchmarks with varying qubit counts distributed over a fixed number of nodes. 
    }
    \label{fig:width_sizes}
\end{figure*}

\begin{figure*}[t]
    \centering
    \begin{subfigure}[t]{0.32\textwidth}
        \centering
        \includegraphics[width=\textwidth]{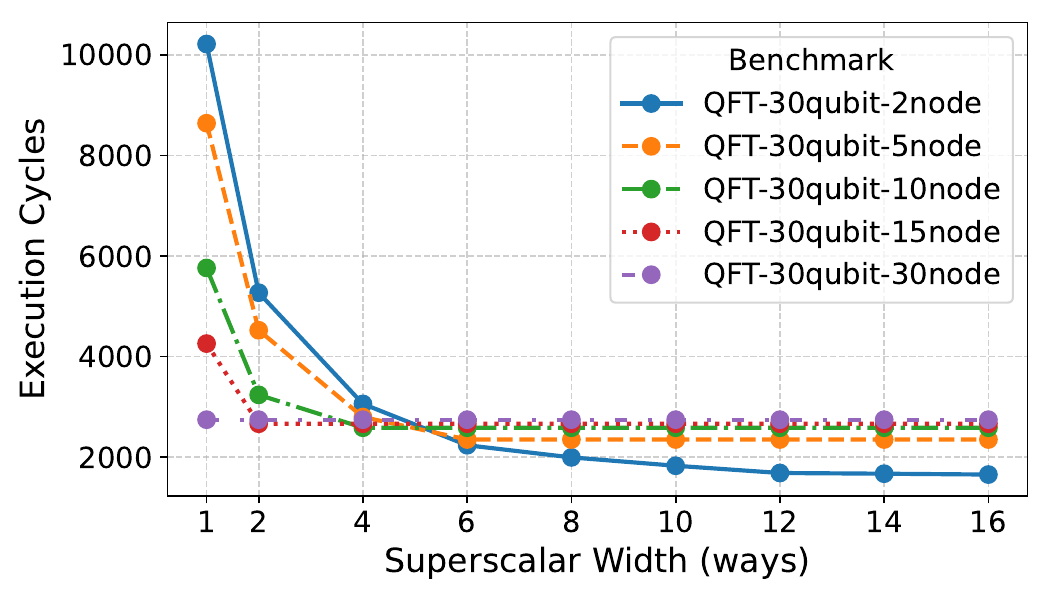}
        \vspace*{-1\baselineskip}
        \caption{QFT}
        \label{fig:qft_width}
    \end{subfigure}
    \begin{subfigure}[t]{0.32\textwidth}
        \centering
        \includegraphics[width=\textwidth]{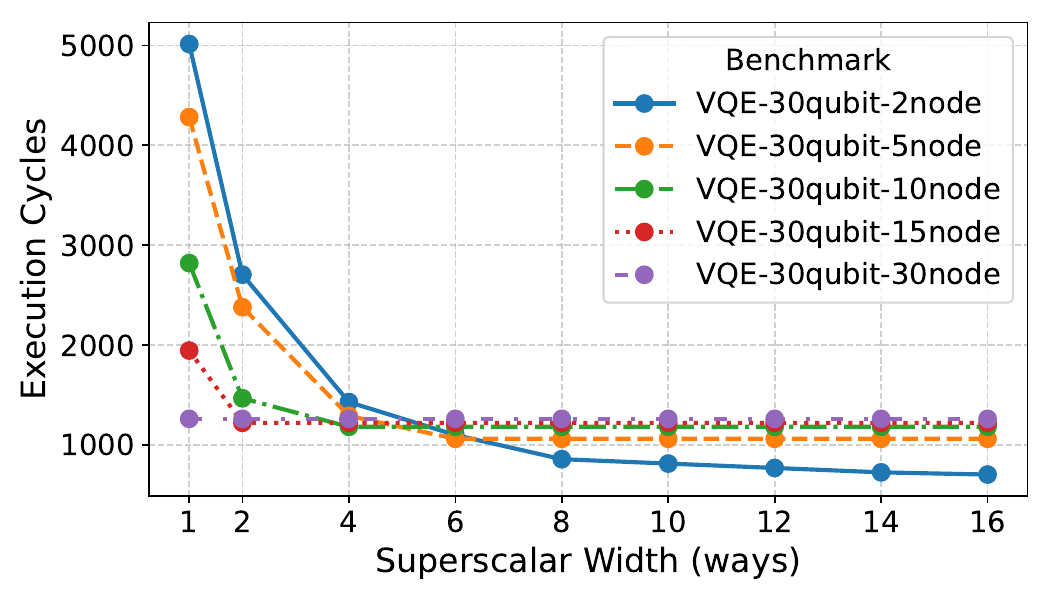}
        \vspace*{-1\baselineskip}
        \caption{VQE}
        \label{fig:vqe_width}
    \end{subfigure}
    \begin{subfigure}[t]{0.32\textwidth}
        \centering
        \includegraphics[width=\textwidth]{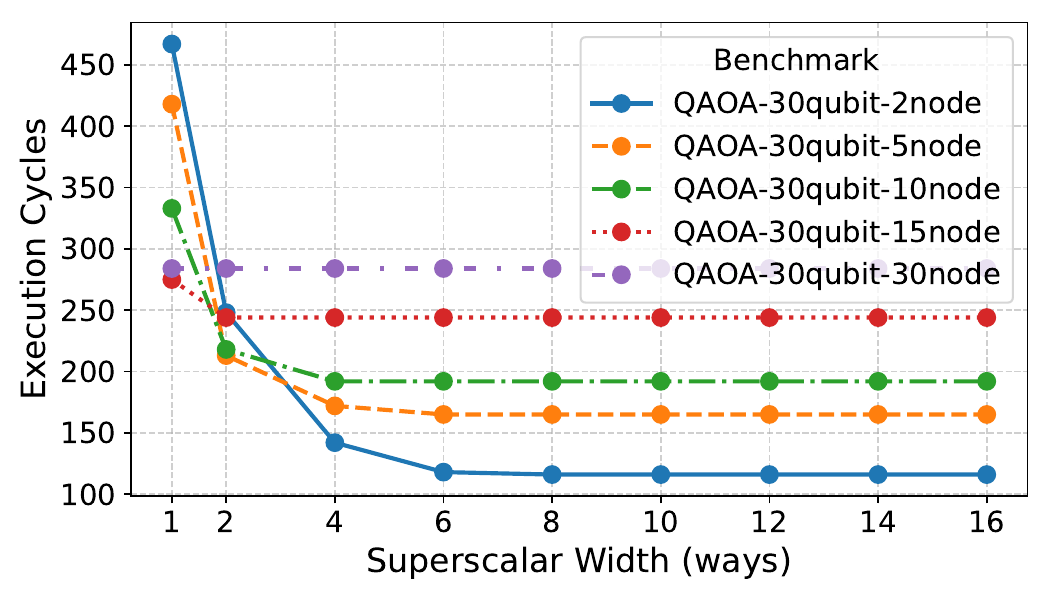}
        \vspace*{-1\baselineskip}
        \caption{QAOA}
        \label{fig:qaoa_width}
    \end{subfigure}
\vspace*{-1\baselineskip}
    \caption{Impact of superscalar width on execution cycles across benchmarks with a fixed number of qubits distributed over varying numbers of nodes. 
    }
    \label{fig:width_nodes}
\end{figure*}
\textbf{Circuit-size Scaling: } In this experiment, we increase the number of qubits on each node by increasing the circuit size while keeping the number of nodes the same. 
For Hamiltonian simulation, GHZ, and VQE with linear entanglement ansatz, the max number of remote gates per node remains constant as the circuit size grows (e.g., from 50 to 150 qubits, the max number of remote gates per node remains 4 for Hamiltonian simulation benchmarks), and these remote operations are strictly serialized. Thus, both scalar and superscalar QNPUs yield negligible gains compared to the monolithic baseline. In contrast, for BV, QFT, VQE with full entanglement ansatz, and QAOA circuits, when the circuit size increases from 50 to 150 qubits, both the total and per-node number of remote gates increase. In these cases, the superscalar QNPU delivers high reductions in execution time (70–75\%), as wider pipelines effectively exploit parallelism across independent remote gates.

\textbf{Node-count Scaling: }
In this experiment, we evaluate scaling of the number of nodes while fixing the circuit size to 150 qubits. For Hamiltonian simulation, GHZ, and VQE with linear entanglement ansatz, increasing the node count does not constantly increase the maximum number of remote gates per node. However, because all remote operations across nodes are serialized, distributing the circuit over more nodes introduces additional inter-node dependencies, which increases the overall execution time. In contrast, for benchmarks such as QFT, VQE with full entanglement ansatz, and QAOA, distributing the same circuit across more nodes increases the total number of remote gates but decreases the maximum number of remote gates assigned to each node. This reduction in per-node communication burden shortens execution time, leading all three architectures to benefit from lower latency as node count increases.

\subsection{Impact of Superscalar Width}
In the previous subsection~\ref{subsec:eva}, we used a fixed 4-way superscalar QNPU. To better understand how the superscalar width influences performance, we extend our study to multiple configurations (2-, 4-, 6-, 8-, 10-, 12-, 14-, and 16-way) and analyze their impact across different benchmarks. We focus on QFT, VQE with full entanglement ansatz, and QAOA, since these workloads contain remote operations that can potentially be executed in parallel. QFT circuits involve all-to-all interactions via controlled-phase gates, making them highly communication-intensive when distributed across nodes. VQE circuits with a full-entanglement ansatz also exhibit communication-intensive behavior. In contrast, our QAOA evaluation targets MaxCut on 2-regular graphs, producing relatively sparse circuits with fewer parallelizable remote operations than QFT and VQE.

\textbf{Circuit-size scaling: }  
Fig.~\ref{fig:width_sizes} shows results for QFT, VQE, and QAOA circuits of 30, 60, and 90 qubits distributed across 5 nodes. With a fixed node count, increasing circuit size introduces more remote operations per node, creating greater opportunities for parallel execution. As a result, wider superscalar pipelines remain effective for larger circuits. For example, the 90-qubit QFT continues to benefit from up to 12- and even 16-way designs, whereas the 30-qubit QFT saturates at 6-way. Across benchmarks, the 90-qubit QAOA saturates around 10-way, while the 90-qubit QFT and VQE still show gains beyond 10-way. This difference arises from QAOA’s relatively sparse structure, which provides fewer independent remote gates to exploit in parallel.

\textbf{Node-count scaling: } 
Fig.~\ref{fig:width_nodes} reports results for 30-qubit QFT, VQE, and QAOA circuits distributed across 2, 5, 10, 15, and 30 nodes. For the fixed QFT-30-qubit circuit, the benefit of larger superscalar width decreases as the node count increases. With 2 nodes, wider designs provide clear speedups; with 5 nodes, performance saturates at 6-way; with 10 nodes, at 4-way; and with 15 nodes, at 2-way. When distributed across 30 nodes, the superscalar design offers no benefit. This trend reflects the fact that distributing a fixed circuit over more nodes reduces the number of remote operations per node, thereby also reducing opportunities of independent remote gates that can be executed in parallel. VQE and QAOA exhibit similar patterns, though they saturate at different superscalar widths.

\section{Conclusions}
In this paper, we present a novel quantum supercomputer architecture featuring decoupled processing units, with QPUs for computation and QNPUs for communication.
We develop the instruction set architecture and microarchitecture for the QNPU to improve communication efficiency particularly for communication-intensive workloads. 
Furthermore, by extending OpenQASM to our proposed DistQASM for simplified programming of distributed quantum algorithms, we have provide a complete solution that bridges the gap between quantum applications and the underlying hardware infrastructure. 
To evaluate our design, we implement a cycle-level performance simulator that evaluates the timing behavior of the proposed architecture under distributed quantum workloads. 
The simulation results demonstrate that these architectural innovations deliver substantial performance improvements compared to conventional monolithic designs.

\section*{Acknowledgements}

This material is based upon work supported by the U.S. Department of Energy, Office of Science, National Quantum Information Science Research Centers, Co-design Center for Quantum Advantage (C2QA) under contract number DE-SC0012704, (Basic Energy Sciences, PNNL FWP 76274). This material is also based upon work supported by the U.S. Department of Energy, Office of Science, Office of
Advanced Scientific Computing Research Advanced Quantum Networks for Scientific Discovery (AQNET-SD) project under contract number DE-AC02-06CH11357. This research used resources of the National Energy Research Scientific Computing Center (NERSC), a U.S. Department of Energy Office of Science User Facility located at Lawrence Berkeley National Laboratory, operated under Contract No. DE-AC02-05CH11231. This research used resources of the Oak Ridge Leadership Computing Facility, which is a DOE Office of Science User Facility supported under Contract DE-AC05-00OR22725. The Pacific Northwest National Laboratory is operated by Battelle for the U.S. Department of Energy under Contract DE-AC05-76RL01830.

The work is also supported in part by NSF grants OSI-2410675, PHY-2325080
(with a subcontract to NC State University from Duke University), OMA-2120757 (with a subcontract to NC State University
from the University of Maryland) and by the U.S. Department
of Energy, Advanced Scientific Computing Research, under
contract number DE-SC0025384.

\bibliographystyle{ACM-Reference-Format}
\bibliography{sample-base}

\end{document}